\documentclass[usenatbib]{mn2e}
\usepackage{graphicx}

\def\cite#1{\citealt{#1}}

\def\inpress{in press}

\def\defjournal#1#2{\def#1{{\rm#2}}}

\defjournal\AcA{Acta Astron.}
\defjournal\ibvs{Inf. Bull. Var. Stars}

\defjournal\aj{AJ}
\defjournal\an{Astron. Nachr.}
\defjournal\apj{ApJ}
\defjournal\apjl{ApJ}
\defjournal\apjs{ApJS}
\defjournal\apss{Ap\&SS}
\defjournal\aap{A\&A}
\defjournal\aapr{A\&A\ Rev.}
\defjournal\aaps{A&AS}
\defjournal\azh{AZh}
\defjournal\mnras{MNRAS}
\defjournal\pasp{PASP}
\defjournal\pasj{PASJ}
\defjournal\sovast{Soviet\ Ast.}
\defjournal\ssr{Space\ Sci.\ Rev.}
\defjournal\nat{Nature}
\defjournal\iaucirc{IAU\ Circ.}

\title{V344 Lyr: an unusual large-amplitude SU UMa-type dwarf nova
       with a short supercycle}

\author[T. Kato et al.]{
       Taichi Kato$^1$,
       Gary Poyner$^2$,
       Timo Kinnunen$^3$ \\
       $^1$ Department of Astronomy, Faculty of Science,
       Kyoto University, Sakyo-ku, Kyoto 606-8502 Japan \\
       $^2$ BAA Variable Star Section,
       67 Ellerton Road, Kingstanding, Birmingham B44 0QE, England \\
       $^3$ Sinirinnantie 16, SF-02660 Espoo, Finland
}

\date{Accepted.
      Received;
      in original form}

\pagerange{\pageref{firstpage}--\pageref{lastpage}}
\pubyear{2001}

\begin{document}

\maketitle

\label{firstpage}

\begin{abstract}
   We studied the large-amplitude SU UMa-type dwarf nova V344 Lyr.
A combination of our observations and reports to VSNET has
yielded an extensive coverage of outbursts between 1994 July and
2001 June.  The analysis of this data showed a mean supercycle length
of 109.6 d.  This value is one of the smallest among known SU UMa-type
dwarf novae (except unusual ER UMa-type dwarf novae).  The outburst
amplitude of V344 Lyr ($\sim$5.5 mag) is found to be much
larger than those (3.5$\pm$0.3 mag) of SU UMa-type dwarf novae with
similar supercycle lengths.  Such a deviation of the amplitude in V344 Lyr
is difficult to explain by the inclination effect.  The extreme outburst
parameters of V344 Lyr would require an additional mechanism to effectively
reduce the quiescent luminosity or to increase the outburst frequency.
\end{abstract}

\begin{keywords}
accretion: accretion disks --- stars: cataclysmic
           --- stars: dwarf novae
           --- stars: individual (V344 Lyr)
\end{keywords}

\section{Introduction}
   Dwarf novae are a class of cataclysmic variables (CVs), which are
close binary systems consisting of a white dwarf and a red dwarf secondary
transferring matter via the Roche lobe overflow.  Thermal instability
in the resultant accretion disk is now widely believed to cause dwarf
nova-type outbursts (see \citet{osa96review} for a review).
In dwarf novae whose binary mass-ratios ($q$=$M_2$/$M_1$) are small
enough, a different kind of instability -- tidal instability -- occurs
\citep{whi88tidal}, which is now widely believed to be a cause of
superoutburst and superhumps in SU UMa-type dwarf novae (for a recent
review of SU UMa-type stars and their observational properties, see
\citet{war95suuma}).

   Among SU UMa-type dwarf novae, some objects show large-amplitude
outbursts.  Although there is evidence that outburst amplitudes of dwarf
novae are known to comprise a continuum from small to large amplitudes
(see a discussion in \cite{pat96alcom}), these large-amplitude objects
are sometimes symbolically known as TOADs (Tremendous Outburst
Amplitude Dwarf Novae: \cite{how95TOAD}).  At the extreme end, there exist
so-called WZ Sge-type dwarf novae (originally proposed by \citet{bai79wzsge};
see also \citet{dow81wzsge} and \citet{odo91wzsge}).  WZ Sge-type dwarf
novae (and related systems) are known to show peculiar outburst
characteristics among SU UMa-type dwarf novae:
1) long interval between outbursts (WZ Sge itself has a recurrence
time of 23 to 33 yr, which is the longest among all known dwarf novae),
2) lack or very low frequency of normal outbursts, which are more abundant
than superoutbursts in usual SU UMa-type dwarf novae (no normal outburst
has ever been observed in WZ Sge itself) and 3) extremely long
(up to $\sim$100 d) and bright superoutbursts.

   V344 Lyr is a dwarf nova discovered by \citet{hof66an289139}.
\citet{kat93v344lyr} detected superhumps with a period of 0.09145(2) d,
which confirms that V344 Lyr is a member of SU UMa-type dwarf novae with
long orbital periods.
The reported range of variability (13.8 -- $[$20 $V$) makes V344 Lyr
a candidate for rare large-amplitude dwarf novae with long orbital periods.
Since many of large-amplitude SU UMa-type dwarf novae are known to be
most infrequently outbursting systems, the proposed supercycle of 240 d
\citep{kat93v344lyr} would make V344 Lyr an exceptional object.
In order to clarify the situation,
we examined our observations and the observations between 1994 July and
2001 June, made by the VSNET Collaboration
\footnote{http://www.kusastro.kyoto-u.ac.jp/vsnet/}.

\section{Observations and results}

\begin{table*}
\caption{Outbursts of V344 Lyr}\label{tab:burst}
\begin{center}
\begin{tabular}{cccccccc}
\hline\hline
JD start & peak mag & length (d) & type &
JD start & peak mag & length (d) & type \\
\hline
2449537 & 14.4 & 13 & super     & 2450651 & 14.3 & $>$8 & super   \\
2449566 & 15.0 &  1 & normal    & 2450674 & 15.7 & 1$^a$ & normal \\
2449604 & 14.0 &  2 & normal    & 2450689 & 15.1 & 1$^a$ & normal \\
2449629 & 14.2 &  4 & normal    & 2450716 & 15.0 & 1$^a$ & normal \\
2449651 & 14.4 &  3 & normal    & 2450727 & 15.0 &  3 & normal    \\
2449662 & 14.8 &  2 & normal    & 2450755 & 15.2 &  2 & normal    \\
2449857 & 14.5 &  4 & normal    & 2450771 & 14.4 & $>$4 & super?  \\
2449867 & 15.8 & 1$^a$ & normal & 2450882 & 14.2 & $>$9 & super   \\
2449923 & 15.0 &  2 & normal    & 2450992 & 14.3 & 10 & super     \\
2449934 & 14.3 &  2 & normal    & 2451042 & 14.7 &  2 & normal    \\
2449946 & 14.3 & 13 & super     & 2451060 & 15.3 & 1$^a$ & normal \\
2450001 & 14.6 &  3 & normal    & 2451094 & 14.2 & 16 & super     \\
2450062 & 14.6 & 1$^a$ & normal & 2451271 & 14.6 &  3 & normal    \\
2450184 & 14.3 & $>$7 & super   & 2451310 & 14.4 & $>$8 & super   \\
2450225 & 15.2 & 1$^a$ & normal & 2451406 & 14.6 &  3 & normal    \\
2450270 & 15.3 &  2 & normal    & 2451416 & 15.5 & 1$^a$ & normal \\
2450290 & 14.6 &  2 & normal    & 2451421 & 14.3 & 16 & super     \\
2450300 & 14.7 &  3 & normal    & 2451465 & 15.1 & 1$^a$ & normal \\
2450312 & 14.7 & $>$3 & super?  & 2451520 & 14.6 & $>$8 & super   \\
2450340 & 15.7 & 1$^a$ & normal & 2451628 & 14.7 & $>$3 & super?  \\
2450370 & 14.6 &  4 & normal    & 2451674 & 15.0 & 1$^a$ & normal \\
2450418 & 14.4 & $>$4 & super?  & 2451737 & 14.6 & 1$^a$ & normal \\
2450506 & 15.5 & 1$^a$ & normal & 2451795 & 15.2 & 1$^a$ & normal \\
2450514 & 15.3 & 1$^a$ & normal & 2451800 & 15.2 &  4 & normal?   \\
2450545 & 14.6 & $>$6 & super?  & 2451815 & 14.5 &  2 & normal?   \\
2450585 & 15.2 & 1$^a$ & normal & 2451865 & 14.6 & 1$^a$ & normal \\
2450607 & 14.9 &  2 & normal    & 2452068 & 14.4 & $>$4 & super   \\
2450639 & 14.6 &  4 & normal    &         &      &      &         \\
\hline
 \multicolumn{8}{l}{$^{a}$ Single observation.} \\
 \multicolumn{8}{l}{$^{b}$ Double peaks.} \\
\end{tabular}
\end{center}
\end{table*}

\begin{figure*}
  \includegraphics[angle=0,height=9cm]{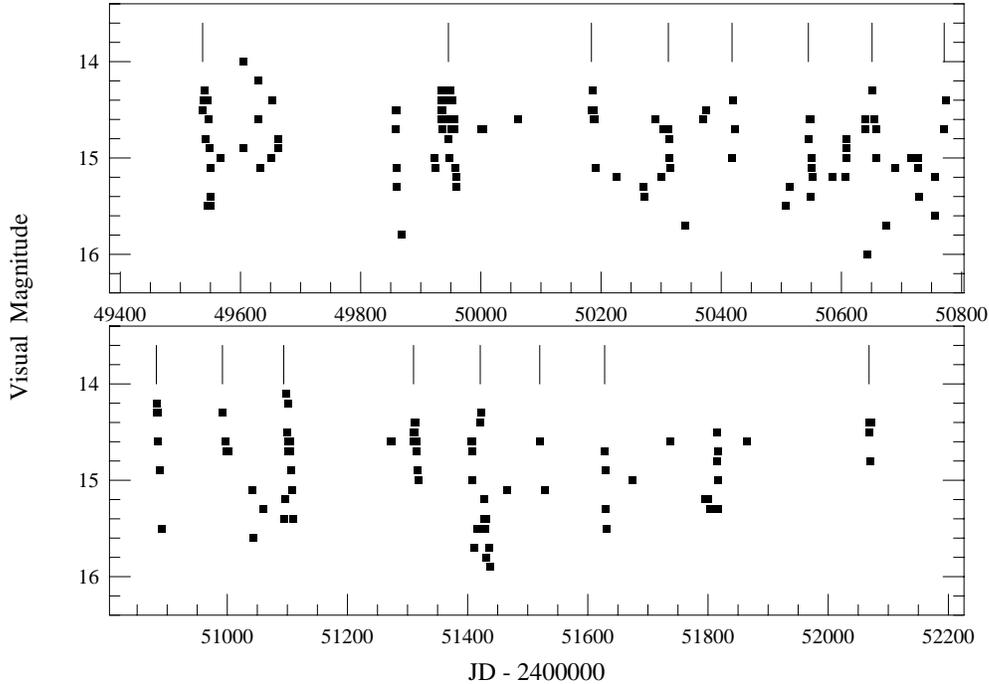}
  \caption{Overall light curve of V344 Lyr.  Superoutbursts
  are marked with ticks.  Upper-limit observations are not plotted for
  simplicity.}
  \label{fig:lc}
\end{figure*}

\subsection{Outburst cycle length and statistics}

   The observations by the authors and those from VSNET were made visually,
using $V$-magnitude calibrated comparison stars.  The typical error of
visual estimates was 0.2 mag, which will not affect the following
discussion.  The total number of observations was 1089.  The number of
positive observations was 209, corresponding to the outburst duty cycle
of 19\%.  Although this value may have suffered from some degree of
selection bias, the large duty cycle is already unusual for a system
with a large outburst amplitude.
We selected outbursts from these data, which are summarized in Table
\ref{tab:burst} and Figure \ref{fig:lc}.

   Most outbursts were unambiguously classified as either
normal outbursts or superoutbursts based on their duration and peak
magnitudes, but the faintness of the object and sometimes unfavourable
observing conditions made some of classifications slightly ambiguous.
Such outbursts are flagged as ``?".

\subsection{Outburst amplitude}

   Since the outburst amplitude is sensitive to outburst and quiescent
magnitudes, we have tried to recalibrate these values using the modern
$V$ scale.  Since recent outburst observations have been made using
modern $V$-magnitude comparison stars, we have adopted $V$=14.0 as the
maximum magnitude. [This value is in agreement with the $V$=14.2 obtained
by \citet{kat93v344lyr}.  Even a better agreement would be achieved when
considering that the CCD observation by \citet{kat93v344lyr} started
a few days after the visual maximum.]

   Regarding the quiescent magnitude, \citet{hof66an289139} stated
that ``the object is invisible on the Palomar Sky Survey plates", from
which an upper limit of 20 has been adopted in the literature.  We have
confirmed this finding by a direct inspection of paper reproductions
of the POSS I plates.  The object is also missing from the USNO A1.0 and
A2.0 catalogs, which reach a limiting magnitude of 19.5.
We have also found that the object is faintly seen on the digitized
POSS I plates, available at the USNOFS Image and Catalogue Archive,
and yielded a magnitude of $R$=19.5$\pm$0.3, by comparison with the modern
comparison stars.  Since short-period dwarf novae have colors close to
$V-R$=0 even in quiescence, this value is considered to be a good
approximation of the quiescent $V$ magnitude.

   We have also inspected the available images, on which V344 Lyr was
probably recorded at or close to its minimum (the object was recorded
in outburst on POSS II plates).  The images taken at Ouda Station, Kyoto
University \citep{Ouda}, on 1991 February 23 have yielded
$I_{\rm c}$=17.8$\pm$0.2.  An unfiltered CCD image on 1996 July 17 shows
no hint of the object down to 19.0 mag.  The Guide Star Catalog plates,
as given in \citet{DownesCVatlas1}, show the object at 18--19 mag.
The above results are summarized in Table \ref{tab:quimag}.
These results may suggest a considerable variation of the quiescent
magnitude, but it would not be surprising that the object may not have
completely reached quiescence in some observations, when one takes the
large outburst duty cycle of 19 \% into account.  From these observations,
we adopted 5.5$\pm$0.3 mag for the outburst amplitude of V344 Lyr.

\begin{table}
\caption{Quiescent magnitudes of V344 Lyr.}\label{tab:quimag}
\begin{center}
\begin{tabular}{lccc}
\hline\hline
Source & Year     & Magnitude & Band \\
\hline
DSS 1      & 1955.555 & 19.5$\pm$0.3 & POSS red \\
GSC plate scan & 1982.390 & 18--19       & $V$ \\
This study & 1991.147 & 17.8$\pm$0.2 & $I_{\rm c}$ \\
This study & 1996.543 & [19.0 & unfiltered CCD$^a$ \\
\hline
 \multicolumn{4}{l}{$^{a}$ System close to $R_{\rm c}$.} \\
\end{tabular}
\end{center}
\end{table}

\section{Discussion}

\subsection{Length of supercycle}\label{sec:supercycle}

\begin{figure*}
  \includegraphics[angle=0,height=6cm]{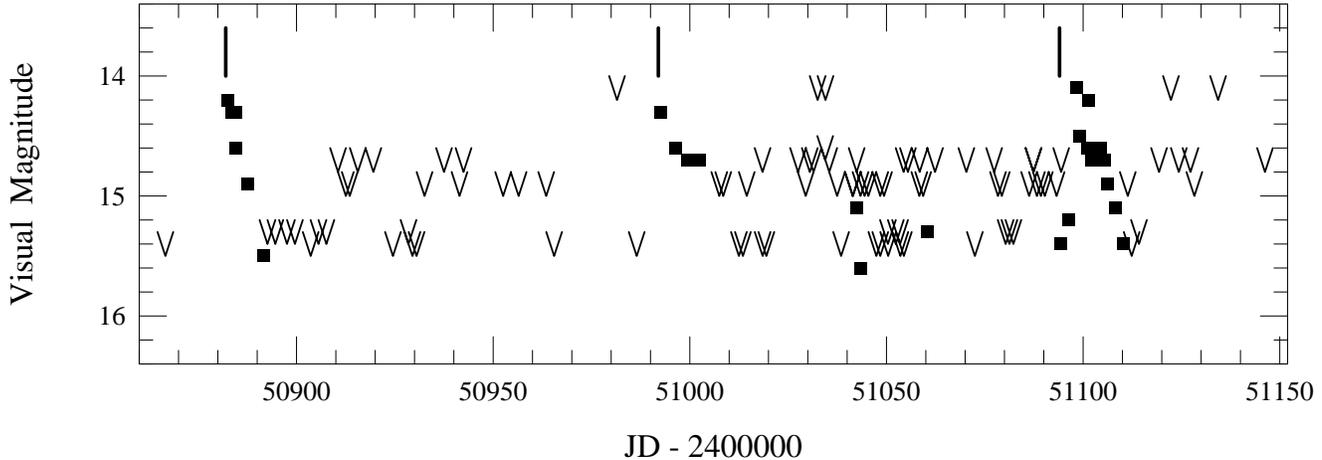}
  \caption{The light curve of V344 Lyr between JD 2450860 (1998 February
  16) and JD 2451152 (1998 December 5).  The ``v" marks represent
  upper-limit observations.  The durations and brightness of the marked
  outbursts qualify them as superoutbursts, confirming the short supercycle
  of $\sim$110 d.  Note that the marks on the superoutbursts correspond to
  the earliest detection of each superoutburst.  The third superoutburst
  was detected during its rise to maximum.  Two short, fainter outbursts
  (normal outbursts) occurring on JD 2451042 and 2451060 were detected
  between the second and third superoutbursts.
  }
  \label{fig:lclarge}
\end{figure*}

   The most prominent sequence of superoutbursts can be found between
JD 2450882 and 2451094 (Figure \ref{fig:lclarge}).  The detection of
three successive superoutbursts establishes the supercycle of $\sim$110 d,
which is half of the estimate by \citet{kat93v344lyr}.
The other superoutbursts and candidate
superoutbursts are well expressed by this period, but the supercycle
does not seem to be extremely stable: the supercycle seems to be
longer ($\sim$120--130 d) in 1994--1995.  Using the best sampled
interval (JD 2450312 -- 2451628), a regression of the observed epochs
of superoutbursts yielded a supercycle of 109.6 d.  This supercycle
length is one of the shortest among known SU UMa-type dwarf novae
(cf. \cite{nog97sxlmi}), except ER UMa stars
(\cite{kat95eruma}; \cite{rob95eruma}; \cite{mis95PGCV}; \cite{nog95rzlmi})
which are unusual SU UMa-type dwarf novae with an exceptionally large
outburst frequency (supercycle lengths are shorter than 60 d; normal
outbursts occur every 3--5 d) and a small outburst amplitude
(for an observational review and theoretical explanations of ER UMa stars,
see \cite{kat99erumareview} and \cite{osa95eruma}, respectively).

\subsection{Outburst amplitude versus supercycle length}

\begin{table}
\caption{Properties of SU UMa-type dwarf novae with short supercycles.}
        \label{tab:supercycle}
\begin{center}
\begin{tabular}{lccccc}
\hline\hline
Name      & $P_{\rm SH}$ (d) &  Max  &  Min  &  Amp$^a$
          & $T_{\rm s}$$^b$ (d) \\
\hline
RZ LMi    & 0.05946 &  14.2  &  17.0  & 2.8 & 19 \\
DI UMa    & 0.0555  &  15.1  &  18.0  & 2.9 & 25 \\
ER UMa    & 0.06573 &  12.9  &  15.8  & 2.9 & 43 \\
V1159 Ori & 0.06861 &  12.8  &  15.4  & 2.6 & 44.6--53.3 \\
IX Dra    & 0.06700 &  15.0  &  17.5  & 2.5 & 45.7--53 \\
SS UMi    & 0.0699  &  13.6  &  16.7  & 3.1 & 84.7 \\
NY Ser    & 0.1064  &  14.53 &  18.2  & 3.7 & 70--100 \\
V503 Cyg  & 0.08101 &  13.95 &  17.5  & 3.5 & 89 \\
V344 Lyr  & 0.09145 &  14.0  &  19.5  & 5.5 & 109.6 \\
FO And    & 0.07411 &  13.9  &  17.5  & 3.6 & 100--140 \\
YZ Cnc    & 0.09204 &  11.0  &  15.0  & 4.0 & 134 \\
V1504 Cyg & 0.0690  &  13.8  &  17.4  & 3.6 & 137 \\
VZ Pyx    & 0.07576 &  11.8  &  15.2  & 3.4 & 152 \\
SU UMa    & 0.0788  &  11.3  &  15.0  & 3.7 & 160 \\
WX Hyi    & 0.07737 &  11.4  &  14.9  & 3.5 & 174 \\
VW Hyi    & 0.07714 &   8.7  &  13.8  & 5.1 & 179 \\
IR Gem    & 0.07094 &  11.4  &  16.3  & 4.9 & 183 \\
V1113 Cyg & 0.0792  &  13.6  &  18.6  & 5.0 & 189.8 \\
TY PsA    & 0.08765 &  11.8  &  15.9  & 4.1 & 202 \\
AY Lyr    & 0.0756  &  12.4  &  18.4  & 6.0 & 210 \\
TT Boo    & 0.07811 &  12.7  &  19.2  & 6.5 & 245 \\
RZ Sge    & 0.07042 &  12.2  &  17.4  & 5.2 & 266 \\
SX LMi    & 0.06850 &  13.3  &  16.8  & 3.4 & 279 \\
Z Cha     & 0.07740 &  12.7  &  15.6  & 2.9 & 287 \\
\hline
 \multicolumn{4}{l}{$^{a}$ Amplitude (mag).} \\
 \multicolumn{4}{l}{$^{b}$ Length of supercycle.} \\
\end{tabular}
\end{center}
\end{table}

   As shown in Section \ref{sec:supercycle}, the supercycle of
V344 Lyr is 109.6 d, which is exceptionally short for an SU UMa-type
dwarf nova with a large outburst amplitude.  Table \ref{tab:supercycle}
lists the properties of SU UMa-type dwarf novae with well-established
short supercycle lengths ($T_{\rm s}<300$ d)\footnote{
  The limit has been chosen for two reasons: (1) $T_{\rm s}$ tends
  to be stable in frequently outbursting systems (i.e. short $T_{\rm s}$
  systems), the extreme cases being ER UMa stars.  Systems with
  superoutbursts less than once per year often do not have a fixed $T_{\rm s}$.
  (2) Due to unavoidable seasonal observational gaps, long $T_{\rm s}$
  systems have uncertainties in unambiguously identifying $T_{\rm s}$.
}.
The source data are from \citet{nog97sxlmi}, with recent revisions and
additions mentioned in the individual notes.  Excluding SX LMi (unusual
low-amplitude system as discussed by \citet{nog97sxlmi}), Z Cha
(high-inclination eclipsing system) at lower right, and V344 Lyr,
we found the following good correlation between $T_{\rm s}$ and outburst
amplitude ($A$) for systems $T_{\rm s}>80$ d (i.e. normal SU UMa stars).

\begin{equation}
A = 2.04 \log(T_{\rm s}) - 0.4. \label{equ:reg1}
\end{equation}

\begin{figure}
  \includegraphics[angle=0,height=6cm]{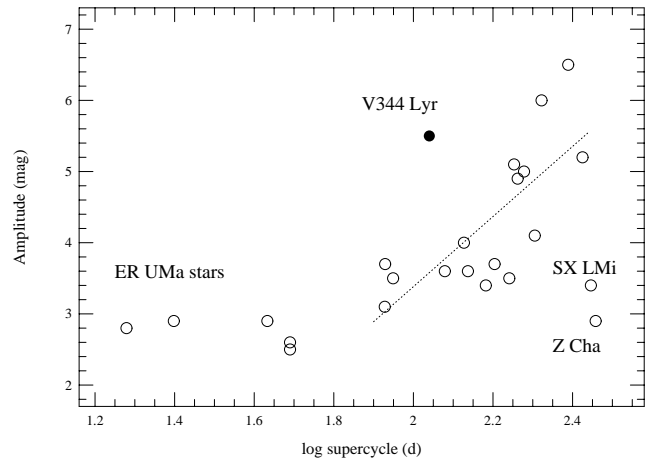}
  \caption{Log $T_{\rm s}$ versus outburst amplitudes drawn from Table
  \ref{tab:supercycle}.  There is a tight relation between $T_{\rm s}$ and
  outburst amplitudes.  The dotted line represents Eq. \ref{equ:reg1}.
  Two systems at significantly below this line include SX LMi and Z Cha,
  a unusual low-amplitude system and a high-inclination eclipsing system,
  respectively.  The deviation of V344 Lyr above this line is evident.
  }
  \label{fig:tsvsamp}
\end{figure}

   Most remarkable is that the outburst amplitudes of SU UMa-type dwarf
novae are confined to a narrow range (3.1 to 4.0 mag) for systems between
$T_{\rm s}$=84 and $T_{\rm s}$=174, except V344 Lyr.  This distribution
has a mean amplitude of 3.5 mag and a standard deviation of 0.3 mag.
The observed amplitude of V344 Lyr (5.5$\pm$0.3 mag) is 7$\pm$1$\sigma$
larger than the average.  Even adopting a conservative upper limit of
the minimum magnitude of 18.5 (from GSC plate scan), the value is still
3$\sigma$ above the average.

   \citet{war87CVabsmag} presented, with an assumption of an optically
thick disk, a formulation of the dependence of the absolute $V$ magnitude
on the system inclination.  A disk seen at nearly edge-on
({\it i}=85$^{\circ}$) is 3.5 mag brighter than the pole-on
({\it i}=0$^{\circ}$) view.  Although this inclination effect explains
the existence of low-amplitude systems among eclipsing systems (e.g. Z Cha),
it would be difficult to explain the existence of high-amplitude systems,
since a pole-on disk is expected to be only 0.4 mag brighter than systems
with moderate inclinations \citep{war87CVabsmag}.  Such a large outburst
amplitude as in V344 Lyr would thus be difficult to reproduce simply by the
geometrical effect.  Since superoutbursts of V344 Lyr have quite
typical properties of usual SU UMa-type dwarf novae \citep{kat93v344lyr},
it is unlikely the superoutbursts of V344 Lyr are intrinsically brighter,
which implies that the quiescent accretion disk of V344 Lyr is fainter
than the average.

   The standard disk instability theory expects that the quiescent
luminosity is a strong function of the mass-transfer rate and the quiescent
viscosity (cf. \cite{war87CVabsmag}; \cite{sma89CVabsmag};
\cite{can98DNabsmag}).  Since the same theory expects that the supercycle
length is inversely proportional to the mass-transfer
rate \citep{ich94cycle}, the mass-transfer rate in V344 Lyr
is expected to be higher than the average, which would work to reduce
the outburst amplitude.  The modification of the quiescent viscosity
could be a viable idea, but is against the observed high frequency of
normal outbursts.  V344 Lyr thus shows a considerable departure from
known SU UMa-type systems, and may require an additional mechanism
to effectively reduce the quiescent luminosity or to increase the
outburst frequency.  Further detailed observations of V344 Lyr, especially
orbital parameters and long-term quiescent photometry are therefore
necessary.

\subsection*{Notes on individual objects in Table \ref{tab:supercycle}:}
\begin{description}

\item[DI UMa] -- Large change of $T_{\rm s}$ has been inferred
  \citep{fri99diuma}.
  The value given in the table corresponds to the most stable outbursting
  period (\cite{kat96diuma}; Kato et al. in preparation).

\item[V1159 Ori] -- Newly determined $T_{\rm s}$ from \citet{kat01v1159ori}.

\item[IX Dra] -- Newly discovered ER UMa-type dwarf nova.  The data are from
  \citet{ish01ixdra}.

\item[SS UMi] -- $T_{\rm s}$ from \citet{kat00ssumi}.  The maximum magnitude of
  $V$=12.6 \citep{mas82ssumi} was probably in error; no such bright
  outburst has been observed in plate searches (see also \cite{ric89ssumi};
  \cite{kat98ssumi}).  The maximum magnitude is taken from the recent
  VSNET observations.

\item[FO And] -- $T_{\rm s}$ and maximum magnitude have been revised using
  VSNET observations.

\item[VZ Pyx] -- $T_{\rm s}$ listed in \citet{kat97vzpyx} and
  \citet{nog97sxlmi}
  was erroneous.  The new $T_{\rm s}$ is determined from VSNET observations,
  which show very regular supercycles, resembling those of YZ Cnc.
  The maximum and minimum magnitudes are also from VSNET observations,
  based on the newly calibrated $V$-magnitude sequence.

\item[SU UMa] -- Typical $T_{\rm s}$ is given.  The system sometimes shows
  reduced outburst and supuroutburst frequencies \citep{ros00suuma}.

\item[WX Hyi] -- Typical $T_{\rm s}$, as measured from VSNET observations
  between 1996 and 1998, is given.  The system is known to sometimes show
  a reduced frequency of superoutbursts.

\item[VW Hyi] -- Revised maximum and minimum magnitudes from VSNET
  observations.

\item[IR Gem] -- Revised $T_{\rm s}$ based on the analysis of the VSOLJ data.
  Revised superhump period from \citet{kat01irgem}.  The maximum
  and minimum magnitudes are from VSNET observations and citet{mis96sequence},
  respectively.

\item[V1113 Cyg] -- Mean $T_{\rm s}$ from \citet{kat01v1113cyg}, which reported
  a variation of supercycles between 169 and 229 d.  \citet{kat01v1113cyg}
  reported rather unusual outburst characteristics, having a low ratio
  of (normal outbursts)/(superoutbursts).  The minimum magnitude given is our
  new calibration of the DSS 1, while \citet{liu99CVspec1} reported a
  rough estimate of $V$=20.8 from their spectroscopic observation.
  The system requires further detailed study in quiescence.

\item[TY PsA] -- Revised $T_{\rm s}$ and maximum magnitude from VSNET
  observations.

\item[AY Lyr] -- Revised $T_{\rm s}$ and maximum magnitude from VSNET
  observations.
  The minimum magnitude is taken from \citet{mis96sequence}.

\item[SX LMi] -- Mean $T_{\rm s}$ from \citet{kat01sxlmi}, which showed a
  variation of supercycles between 250 and 312 d.

\end{description}

\vskip 2mm

The following objects in \citet{nog97sxlmi} have been excluded from the list:

\begin{description}

\item[CI UMa] -- \citet{nog97ciuma} listed a possible supercycle of 140 d,
  while recent reports to VSNET show that the intervals are typically longer
  and outbursts occur rather irregularly (see also \citet{kat00ssumi}).

\item[V630 Cyg] --  \citet{nog97sxlmi} listed a possible supercycle of 290 d.
  Recent study by \citet{nog01v630cyg} indicates that supercycles vary more
  irregularly.

\end{description}

\section{Summary}

   We observed the large-amplitude SU UMa-type dwarf nova V344 Lyr.
Combined with reports to VSNET, we have succeeded in determining
its mean supercycle length as 109.6 d.  This value is one of the
smallest among known SU UMa-type dwarf novae (except unusual ER UMa-type
dwarf novae).  The observed outburst amplitude (5.5$\pm$0.3 mag)
in V344 Lyr is found to be exceptionally large for a system with
such a short supercycle.  Since a short supercycle strongly suggests
a relatively high mass-transfer rate, the extreme outburst parameters
of V344 Lyr would require a mechanism to effectively reduce the quiescent
luminosity or to increase the outburst frequency.

\section*{Acknowledgments}

The authors are grateful to VSNET members, especially to L. Szentasko,
T. Vanmunster, J. Pietz, L. T. Jensen, E. Broens, M. Reszelski
for providing vital observations.  The authors are grateful to the staff
of the VSOLJ (Variable Star Observers League) for their on-line database.
We are also grateful to an anonymous referee, whose comments significantly
improved the paper.
This research has made use of the USNOFS Image and Catalogue Archive
operated by the United States Naval Observatory, Flagstaff Station
(http://www.nofs.navy.mil/data/fchpix/).

\label{lastpage}

\end{document}